\documentclass[11pt]{article}
\usepackage{authblk}
\usepackage[colorlinks,linkcolor=blue,anchorcolor=red,citecolor=red,pdfstartview=FitH]{hyperref}
\usepackage{graphics,graphicx}
\usepackage{subfig}
\usepackage{mathtools}
\usepackage{amsmath,amssymb,amsthm,accents,amsfonts}
\usepackage{color}
\usepackage{cite}
\usepackage{fullpage}
\usepackage{caption}
\usepackage{hyperref}\hypersetup{colorlinks=true,linkcolor=blue,filecolor=magenta,urlcolor=blue,}
\DeclareMathAlphabet{\mathpzc}{OT1}{pzc}{m}{it}

\DeclareMathOperator{\sech}{sech}

\numberwithin{equation}{section}
\allowdisplaybreaks
\begin{document}

\title{\bf {Darboux Transformation for the Hirota equation}}
\author[1,2,3]{\bf Halis Yilmaz\footnote{E-mail: Halis.Yilmaz@glasgow.ac.uk, halisyilmaz@dicle.edu.tr}}

\affil[1]{School of Mathematics and Statistics, University of Glasgow, Glasgow G12 8QQ, UK}
\affil[2]{Department of Mathematics, Mimar Sinan Fine Arts University, Istanbul, Turkey}
\affil[3]{Department of Mathematics, University of Dicle, 21280 Diyarbakir, Turkey}
\date{}
\maketitle

\begin{abstract}
The Hirota equation is an integrable higher order nonlinear Schr\"{o}dinger type equation which describes the propagation of ultrashort light pulses in optical fibers. We present a standard Darboux transformation for the Hirota equation and then construct its quasideterminant solutions. The multisoliton and breather solutions of the Hirota equation are given explicitly.
\end{abstract}

\quad{\text{\it{Keywords:}} Hirota equation; Darboux transformation; Quasideterminants.}
\vspace{2mm}

\quad{\text{2020 Mathematics Subject Classification:} 35C08, 35Q55, 37K10}

\section{Introduction}
There exists a large class of nonlinear evolution equations which can be solved analytically. Such equations are called integrable.  Integrable equations constitute an important part of the nonlinear wave theory. The simplest integrable equation which describes the dynamics of deep-water gravity waves is the nonlinear Schr\"odinger (NLS) equation 
\begin{eqnarray}\label{NLS}
  iq_t+q_{xx}+2|q|^2q=0.
 \end{eqnarray}
In 1967, it was first discussed in the general context of nonlinear dispersive waves by Benney and Newell \cite{BN}. In 1968, this equation was also derived by Zakharov in his study of modulational stability of deep water waves \cite{Z68}. In 1972, Zakharov and Shabat found that the NLS equation had a Lax pair and could be solved by the inverse scattering transform (IST) method \cite{ZS}. This equation plays an important role in different physical systems as wide as plasma physics \cite{Z72}, water waves \cite{BN, BR, Z68}, and nonlinear optics \cite{HT1, HT2}. One of the most interesting applications of the NLS equation is that it can be employed to model for short soliton pulses in optical fibres \cite{KA}. However, as the pulses get shorter, various additional effects become important and the NLS model is no longer appropriate. In order to understand these additional effects, Kodama and Hasegawa \cite{K85, KH87} suggested a higher-order NLS equation
\begin{eqnarray}\label{KHeqn}
 iq_{t}+\alpha_{1}q_{xx}+\alpha_{2}|q|^{2}q+i\beta\left[\gamma_{1}q_{xxx}+\gamma_{2}|q|^{2}q_{x}+\gamma_{3}q \left(|q|^{2}\right)_{x}\right]=0,
\end{eqnarray} 
where the $\alpha_{i}$, $\gamma_{i}$ are real constants, $\beta$ is a real spectral parameter and $q$ is a complex-valued function of $x$ and $t$. By choosing $\beta=0$ and $\alpha_{2}=2\alpha_{1}=2$ in this equation, we can easily see that the first three terms form the standard NLS equation \eqref{NLS}.
Generally, the Kodama-Hasegawa higher-order NLS equation \eqref{KHeqn} may not be completely integrable if some restrictions are not imposed on the real constants $\gamma_{i}$ $(i=1,2,3)$. 
Until now it is known that, besides the NLS equation \eqref{NLS} itself, there are four cases in which integrability can be proved via the IST. These are the Chen-Lee-Liu \cite{CLL} derivative NLS equation $(\gamma_{1}:\gamma_{2}:\gamma_{3}=0:1:0)$, the Kaup-Newell \cite{KN} derivative NLS equation $(\gamma_{1}:\gamma_{2}:\gamma_{3}=0:1:1)$, the Hirota \cite{H73} NLS equation $(\gamma_{1}:\gamma_{2}:\gamma_{3}=1:6:0)$ and the Sasa-Satsuma \cite{SS} NLS equation $(\gamma_{1}:\gamma_{2}:\gamma_{3}=1:6:3)$.

In this paper, we consider the Hirota \cite{H73} NLS equation
\begin{eqnarray}\label{HE}
 iq_t+\alpha\left(q_{xx}+2|q|^2q\right)+i\beta\left(q_{xxx}+6|q|^2q_x\right)=0, ~~\alpha, \beta\in \mathbb{R},
\end{eqnarray}
in which $\alpha_{2}=2\alpha_{1}=2\alpha$. This equation is commonly known as the Hirota equation (HE), and we will denote it as such from now on. The HE \eqref{HE} can be used to describe the wave propagation of ultrashort light pulses in optical fibers \cite{K85,KH87,MTC,Agrawal,MM19,YD}. 
It is very interesting to see that the Hirota equation \eqref{HE}  is the sum of the NLS \eqref{NLS} equation ($\alpha=1, \beta=0$) and the complex version of the modified Korteweg-de Vries (mKdV) equation ($\alpha=0, \beta=1$) 
\begin{eqnarray}\label{cmKdV}
   q_t+q_{xxx}+6|q|^2q_x=0
\end{eqnarray}
which is completely integrable \cite{W72,H73}. In the resent years, there has been some interest in solutions of the HE \eqref{HE} obtained by $\it{Darboux-type}$ transformations \cite{Tao,ASN,L13}. These solutions are often written in terms of determinants. These methods have been used to prove various solutions of the HE \eqref{HE} such as multisolitons, breathers and rogue waves.

In 1882, the French mathematician Jean Gaston Darboux \cite{Darboux} introduced a method to solve the Sturm-Louville equation, which is called Darboux transformation (DT) afterwards. Almost a century later, in 1979,  Matveev \cite{Matveev}  realised that the method given by Darboux for the spectral problem of second order ordinary differential equations can be extended to some important soliton equations. Darboux transformations are one of important tools in studying integrable systems. They provide a universal algorithmic procedure to derive exact solutions of integrable systems. 

In the present article, we construct for the first time a standard Darboux transformation for the Hirota equation \eqref{HE}. We underline that the method we use here is based on Darboux's \cite{Darboux} and Matveev's original ideas \cite{Matveev, MS}. Therefore, our approach should be considered on its own merits. Furthermore, our solutions for the HE are written in terms of quasideterminants \cite{Gelfand91,Gelfand05} rather than determinants. It has been proved that quasideterminants are very useful for constructing exact solutions of integrable equations \cite{CJ,GHHN,LN,NY15,R19,WLL,ZWM}, enabling these solutions to be expressed in a simple and compact form.

This paper is structured as follows. In Section \ref{qdet} below, we give a brief review of quasideterminants. In Section \ref{HEqn}, we establish a $2\times2$ eigenfunction and corresponding constant $2\times2$ square matrix for the eigenvalue problems of the Hirota equation \eqref{HE} using two symmetries of the Lax pair of the HE. In Section \ref{DTs}, we state a standard Darboux theorem for the Hirota system. We review the reduced DTs for the HE, which can be considered as a dimensional reduction from $(2+1)$ to $(1+1)$ dimensions. In Section \ref{SHE}, we present the quasideterminant solutions for the HE constructed by the DT. In Section \ref{ParS}, the multisoliton and breather solutions of the Hirota equation are given for both zero and non-zero seed solutions as particular solutions of the HE. The conclusion is given in the final Section \ref{Con}.

\subsection{Quasideterminants}\label{qdet}

In this short section we will list some of the key elementary properties of quasideterminants used in the paper. The reader is referred to the original papers \cite{Gelfand91,Gelfand05} for a more detailed and general treatment. 

Let $M=(m_{ij})$ be an $n\times n$ matrix  with entries over a ring (noncommutative, in general)  has $n^2$ quasideterminants written as $|M|_{ij}$ for $i,j=1,\dots,n$. They are defined recursively by
\begin{equation}\label{expan}
 |M|_{ij}=m_{ij}-r_i^j\left(M^{ij}\right)^{-1}c_j^i ,
\end{equation}
where $r_i^j$ represents the row vector obtained from $i^{th}$ row of $M$ with the $j^{th}$ element removed, $c_j^i$ is the column vector obtained from $j^{th}$ column of $M$ with the $i^{th}$ element removed and $M^{ij}$ is the $(n-1)\times (n-1)$ submatrix obtained by deleting the $i^{th}$ row and the $j^{th}$ column from $M$. Quasideterminants can be also denoted as shown below by boxing the entry about which the expansion is made
\begin{align}
 |M|_{ij}=
         {\left|\begin{array}{cc}
          M^{ij}& c_j^i\\ r_i^j & \boxed{m_{ij}}
         \end{array}\right|}.
\end{align}
If the entries in $M$ commute, then the quasideterminant $|M|_{ij}$ can be expressed as a ratio of determinants
\begin{equation}\label{comdet}
 |M|_{ij}=(-1)^{i+j}\frac{\det M}{\det M^{ij}}.
\end{equation}

\section{Hirota equation}\label{HEqn}
Let us consider the couple Hirota equations
\begin{eqnarray}
    q_t-i\alpha\left(q_{xx}+2q^2r\right)+\beta\left(q_{xxx}+6qrq_x\right)&=&0, \label{HE1}\\
    r_t+i\alpha\left(r_{xx}+2qr^2\right)+\beta\left(r_{xxx}+6qrr_x\right)&=&0, \label{HE2}
\end{eqnarray}
where $q=q(x,t)$ and $r=r(x,t)$ are complex valued functions. Equations \eqref{HE1} and \eqref{HE2} reduce to the Hirota equation \eqref{HE} when $r=q^*$. Here the asterisk superscript on $q$ denotes the complex conjugate.

The Lax pair \cite{Tao} for the couple Hirota equations \eqref{HE1}-\eqref{HE2} is given by
\begin{eqnarray}
 L&=&\partial_x+J\lambda-R, \label{LaxL}\\
 M&=&\partial_t+4\beta J\lambda^3+2U\lambda^2-2V\lambda-W,\label{LaxM}
\end{eqnarray}
where $J$, $R$, $U$, $V$ and $W$ are $2\times 2$ matrices
\begin{eqnarray}\label{JRU}
  J={\left(\begin{array}{cc}
          i & 0\\ 0 & -i
  \end{array}\right)},\hspace{0.5cm}
   R={\left(\begin{array}{cc}
          0& q\\ -r & 0
  \end{array}\right)},\hspace{0.5cm}
   U={\left(\begin{array}{cc}
          i\alpha& -2\beta q\\ 2\beta r& -i\alpha
  \end{array}\right)},
  \end{eqnarray}
 \begin{eqnarray}\label{VW}
  V=\left(\begin{array}{cc}
          i\beta qr & \alpha q+i\beta q_x \\  -\alpha r+i\beta r_x& -i\beta qr
 \end{array}\right),
  W=\left(\begin{array}{cc}
       i\alpha qr+\beta\left(qr_x-rq_x\right)& i\alpha q_x-\beta\left(q_{xx}+2q^2r\right)\\
       i\alpha r_x+\beta\left(r_{xx}+2qr^2\right) & - i\alpha qr-\beta\left(qr_x-rq_x\right)  
 \end{array}\right).\hspace{0.2cm}
\end{eqnarray}
Here $\lambda$ is a spectral parameter. It can be seen that the potential matrix $R$ in \eqref{JRU} has two symmetry properties. One is that it is skew-Hermitian: $R^\dag=-R$. The other one is that $SRS^{-1}=R^*$, where 
\begin{eqnarray}
  S={\left(\begin{array}{cc}
          0 & 1\\ -1 & 0
  \end{array}\right)}.
\end{eqnarray}
Let $\phi=(\varphi, \psi)^{T}$ be a vector eigenfunction for \eqref{LaxL}-\eqref{LaxM} for eigenvalue $\lambda$ so that $L_{\lambda}(\phi)=M_{\lambda}(\phi)=0$. Using the second symmetry, it may be seen that $\tilde{\phi}=S\phi=(\psi^*, -\varphi^*)^{T}$ is another eigenfunction for eigenvalue $\lambda^*$ such that $L_{\lambda^*}(\tilde{\phi})=M_{\lambda^*}(\tilde{\phi})=0$. Using these vector eigenfunctions we can define a square $2\times 2$ matrix eigenfunction $\theta$ with $2\times 2$ eigenvalue $\Lambda$
\begin{eqnarray}\label{thetaLambda}
 \theta=\left(\begin{array}{cc}
        \varphi & \psi^*\\ \psi &- \varphi^*
  \end{array}\right), \hspace{0.3cm}
  \Lambda=\left(\begin{array}{cc}\lambda & 0\\ 0 & \lambda^* \end{array}\right),
\end{eqnarray}
satisfying
\begin{eqnarray}
  &&\theta_{x}+J\theta \Lambda-R\theta=0,\label{LaxEqnX}\\
 && \theta_{t}+4\beta J\theta \Lambda^3+2U\theta \Lambda^2-2V\theta\Lambda-W\theta=0.\label{LaxEqnT}
\end{eqnarray} 

\section{Darboux transformations and Dimensional reductions}\label{DTs}
\subsection{Darboux transformation}
Let us consider the linear operators
\begin{equation}\label{LM}
 L=\partial_x+\sum_{i=0}^nu_i\partial_y^i,\hspace{0.5cm}M=\partial_t+\sum_{i=0}^nv_i\partial_y^i,
\end{equation}
where $u_{i}, v_{i}$  are $m\times m$ matrices.
The standard approach to Darboux transformations \cite{Darboux, Matveev, MS} involves a gauge operator $G_{\theta}=\theta\partial_{y}\theta^{-1}$, where 
$\theta=\theta(x,y,t)$ is an invertible $m\times m $ matrix solution to a linear system 
\begin{eqnarray}
 L(\phi)=M(\phi)=0.
\end{eqnarray}
If $\phi$ is any eigenfunction of $L$ and $M$ then $\tilde{\phi}=G_{\theta}(\phi)$ satisfies the transformed system
\begin{eqnarray}
 \tilde{L}(\tilde{\phi})=\tilde{M}(\tilde{\phi})=0,
\end{eqnarray}
where the linear operators $\tilde{L}=G_{\theta}LG_{\theta}^{-1}$ and $ \tilde{M}=G_{\theta}MG_{\theta}^{-1}$ have the same forms as $L$ and $M$:
\begin{equation}
 \tilde{L}=\partial_x+\sum_{i=0}^n\tilde{u}_i\partial_y^i,\hspace{0.5cm}\tilde{M}=\partial_t+\sum_{i=0}^n\tilde{v}_i\partial_y^i.
\end{equation}

\subsection{Dimensional reduction of Darboux transformation}\label{DRDT}
Here, we describe a reduction of the Darboux transformation from $(2+1)$ to $(1+1)$ dimensions. We choose to  eliminate the $y$-dependence by employing a `separation of variables' technique. The reader is referred to the paper \cite{NGO} for a more detailed treatment. We make the ansatz
\begin{eqnarray}
 \phi &=&\phi^r(x,t)e^{\lambda y},\\
 \theta &=&\theta^r(x,t)e^{\Lambda y},
\end{eqnarray}
where $\lambda$ is a constant scalar and $\Lambda$ an $N \times N$ constant matrix and
the superscript $r$ denotes reduced functions, independent of $y$.
Hence in the dimensional reduction we obtain $\partial_y^{i}\left(\phi\right)=\lambda^i\phi$ and
$\partial_y^{i}\left(\theta \right)=\theta \Lambda^i$ and so the operator $L$ and Darboux transformation $G$ become
 \begin{eqnarray}\label{redL}
  L^r&=&\partial_x+\sum_{i=0}^n u_i\lambda^i,\\
  G^r&=&\lambda-\theta^r\Lambda(\theta^{r})^{-1},
 \end{eqnarray}
where $\theta^r$ is a matrix eigenfunction of $L^r$ such that $L^r\left(\theta^r\right)=0$, with $\lambda$ replaced by the matrix $\Lambda$, that is,
\begin{equation}
	\theta^r_x+\sum_{i=0}^n u_i\theta^r\Lambda^i=0.
\end{equation}
Below we omit the superscript $r$ for ease of notation.

\subsection{Iteration of reduced Darboux Transformations}
In this section we shall consider iteration of the Darboux transformation and find closed form expressions for these in terms of quasideterminants. 

Let $L$ be an operator, form invariant under the reduced Darboux transformation $G_{\theta}=\lambda-\theta \Lambda \theta^{-1}$ discussed above.

Let $\phi=\phi(x,t)$ be a general eigenfunction of $L$ such that $L(\phi)=0$. Then
\begin{eqnarray*}
\tilde{\phi}&=&G_\theta\left(\phi\right)\\
             &=&\lambda\phi-\theta\Lambda\theta^{-1}\phi\\
             &=&\left|\begin{array}{cc} \theta & \phi\\ \theta\Lambda & \boxed{\lambda\phi}\end{array}\right|
\end{eqnarray*}
is an eigenfunction of $\tilde{L}=G_\theta L G_\theta^{-1}$ so that
$\tilde{L}(\tilde{\phi})=\lambda\tilde{\phi}$. Let $\theta_i$ for $i=1,\ldots,n,$ be a particular set
of invertible eigenfunctions of $L$ so that $L(\theta_i)=0$ for $\lambda=\Lambda_i$, and introduce the notation $\Theta=(\theta_1,\ldots,\theta_n)$. To apply the Darboux transformation a second time, let $\theta_{[1]}=\theta_1$ and $\phi_{[1]}=\phi$ be a general eigenfunction of $L_{[1]}=L$.
Then
$\phi_{[2]}=G_{\theta_{[1]}}\left(\phi_{[1]}\right)$ and $\theta_{[2]}=\phi_{[2]}|_{\phi\rightarrow \theta_2}$ are eigenfunctions for $L_{[2]}=G_{\theta_{[1]}}L_{[1]}G_{\theta_{[1]}}^{-1}$.

In general, for $n\geq 1$, we define the $n$th Darboux transform of $\phi$ by
\begin{equation}
 \phi_{[n+1]}=\lambda\phi_{[n]}-\theta_{[n]}\Lambda_n\theta_{[n]}^{-1}\phi_{[n]},
\end{equation}
in which
\begin{equation*}
 \theta_{[k]}=\phi_{[k]}|_{\phi\rightarrow\theta_k}~.
\end{equation*}
For example,
\begin{eqnarray*}
\phi_{[2]}&=&\lambda\phi-\theta_1\Lambda_1\theta_1^{-1}\phi
        =\left|\begin{array}{cc} \theta_1 & \phi\\ \theta_1\Lambda_1& \boxed{\lambda\phi}\end{array}\right|,\\
\phi_{[3]}&=&\lambda\phi_{[2]}-\theta_{[2]}\Lambda_2\theta_{[2]}^{-1}\phi_{[2]}\\
          &=&\left|\begin{array}{ccc} \theta_1 & \theta_2 & \phi\\
            \theta_1\Lambda_1 & \theta_2 \Lambda_2 & \lambda \phi\\
            \theta_1\Lambda_1^2 &\theta_2\Lambda_2^2 & \boxed{\lambda^2\phi} \end{array}\right|.
\end{eqnarray*}
After $n$ iterations, we get
\begin{eqnarray}
\phi_{[n+1]}=\left|\begin{array}{ccccc}
      \theta_1 & \theta_2 \hspace{0.2cm} \ldots \hspace{0.2cm} \theta_n & \phi\\
      \theta_1\Lambda_1 & \theta_2\Lambda_2 \ldots \theta_n\Lambda_n & \lambda \phi\\
      \theta_1\Lambda_1^2 & \theta_2\Lambda_2^2 \ldots \theta_n\Lambda_n^2 & \lambda^2 \phi\\
      \vdots & \vdots \hspace{0.3cm} \ldots \hspace{0.3cm} \vdots & \vdots \\
      \theta_1\Lambda_1^n & \theta_2\Lambda_2^n \ldots \theta_n\Lambda_n^n & \boxed{\lambda^n\phi}\\
     \end{array}
 \right|.
\end{eqnarray}

\section{Constructing Solutions for Hirota Equation}\label{SHE}
In this section we determine the specific effect of the Darboux transformation $G_{\theta}=\lambda-\theta \Lambda \theta^{-1}$ on the operator $L$ given by \eqref{LaxL}. Corresponding results hold for the operator $M$ given by \eqref{LaxM}.  Here the eigenfunction $\theta$ is the solution of the linear system \eqref{LaxEqnX}-\eqref{LaxEqnT} is given explicitly with the eigenvalue $\Lambda$ in \eqref{thetaLambda}. From $\tilde{L}G_{\theta}=G_{\theta}L$, the operator $ L=\partial_x+J\lambda-R$ is transformed to a new operator $\tilde{L}$ in which $J$ is unchanged and
\begin{eqnarray}\label{EqnR}
 \tilde{R}=R-\left[J,\theta \Lambda \theta^{-1}\right].
\end{eqnarray}
For notational convenience, we introduce a $2 \times 2$ matrix $Q$ such that $R=[J,Q]$, and hence
\begin{eqnarray}
  Q=\frac{1}{2i}\left(\begin{array}{cc}& q\\ r &  \end{array}\right),
\end{eqnarray}
where the entries left blank are arbitrary and do not contribute to $R$. From \eqref{EqnR} it follows that 
\begin{eqnarray}\label{Q1}
 \tilde{Q}=Q-\theta \Lambda \theta^{-1}
\end{eqnarray}
which can be written in a quasideterminant structure as
\begin{eqnarray}
 \tilde{Q}=Q+\left|\begin{array}{cc} \theta & I_2\\ \theta\Lambda & \boxed{0_2}\end{array}\right|.
\end{eqnarray}
We rewrite \eqref{Q1} as
\begin{eqnarray}
 Q_{[2]}=Q_{[1]}-\theta_{[1]}\Lambda_1 \theta_{[1]}^{-1}
\end{eqnarray}
where $Q_{[1]}=Q$, $Q_{[2]}=\tilde{Q}$, $\theta_{[1]}=\theta_1=\theta$ and $\Lambda_1=\Lambda$.
Then after $n$ repeated Darboux transformations, we have
\begin{eqnarray}
 Q_{[n+1]}=Q_{[n]}-\theta_{[n]} \Lambda_n \theta_{[n]}^{-1}
\end{eqnarray}
in which $\theta_{[k]}=\phi_{[k]}\mid_{\phi\rightarrow \theta_k}$. We express $P_{[n+1]}$ in quasideterminant form as
\begin{eqnarray}\label{Qn1}
 Q_{[n+1]}=Q+\left|\begin{array}{ccccc}
      \theta_1 & \theta_2 \hspace{0.2cm} \ldots \hspace{0.2cm} \theta_n & 0_2\\
      \theta_1\Lambda_1 & \theta_2\Lambda_2 \ldots \theta_n\Lambda_n & 0_2\\
      \vdots & \vdots \hspace{0.3cm} \ldots \hspace{0.3cm} \vdots & \vdots \\
      \theta_1\Lambda_1^{n-2} & \theta_2\Lambda_2^{n-2} \ldots \theta_n\Lambda_n^{n-2} & 0_2\\
      \theta_1\Lambda_1^{n-1} & \theta_2\Lambda_2^{n-1} \ldots \theta_n\Lambda_n^{n-1} & I_2\\
      \theta_1\Lambda_1^n & \theta_2\Lambda_2^n \ldots \theta_n\Lambda_n^n & \boxed{0_2}\\
     \end{array}
 \right|,
\end{eqnarray}
where each $\theta_i, \Lambda_i$ as a $2 \times 2$ matrix
\begin{eqnarray}\label{thetaLambdai}
 \theta_i=\left(\begin{array}{cc}
        \varphi_i & \psi_{i}^*\\ \psi_i &- \varphi_{i}^*
  \end{array}\right), \hspace{0.3cm}
  \Lambda_i=\left(\begin{array}{cc}\lambda_i & 0\\ 0 & \lambda_{i}^* \end{array}\right)
\end{eqnarray}
in which $i=1, ... , n$. Now let $\Theta^{(n)}$ be a $2\times 2n$ matrix such that
\begin{eqnarray}
 \Theta^{(n)}=\left(\theta_1\Lambda_1^n,\ldots,\theta_n\Lambda_n^n\right)
       =\left(\begin{array}{c}\varphi^{(n)}\\\psi^{(n)} \end{array}\right),
\end{eqnarray}
where
\begin{eqnarray*}
\varphi^{(n)}&=&\left(\lambda_1^n\varphi_1,\lambda_1^{*n}\psi_1^* , \ldots, \lambda_n^n\varphi_n, \lambda_n^{*n}\psi_n^*\right),\\
\psi^{(n)}&=&\left(\lambda_1^n\psi_1,-\lambda_1^{*n}\varphi_1^* , \ldots, \lambda_n^n\psi_n, -\lambda_n^{*n}\varphi_n^*\right)
\end{eqnarray*}
denote $1 \times 2n$ row vectors. Thus, \eqref{Qn1} can be written as
\begin{eqnarray}
  Q_{[n+1]}=Q+\left|\begin{array}{cc} \widehat{\Theta} & E\\ \Theta^{(n)} & \boxed{0_2}\end{array}\right|,
\end{eqnarray}
where $\widehat{\Theta}=\left(\theta_i\Lambda_i^{j-1}\right)_{i,j=1,\ldots,n}$ and $E=\left(e_{2n-1}, e_{2n}\right)$ denote
$2n \times 2n$ and $2n \times 2$ matrices respectively, in which $e_i$ represents a column vector with $1$ in the $i^{th}$ row and zeros elsewhere. Hence, we obtain
\begin{eqnarray}
 Q_{[n+1]}=Q+\left(\begin{array}{cc}
 \left|\begin{array}{cc} \widehat{\Theta} & e_{2n-1}\\  \varphi^{(n)} & \boxed{0}\end{array}\right| &
 \left|\begin{array}{cc} \widehat{\Theta} & e_{2n}\\  \varphi^{(n)} & \boxed{0}\end{array}\right|\\\\
 \left|\begin{array}{cc} \widehat{\Theta} & e_{2n-1}\\  \psi^{(n)} & \boxed{0}\end{array}\right| &
 \left|\begin{array}{cc} \widehat{\Theta} & e_{2n}\\  \psi^{(n)} & \boxed{0}\end{array}\right| \end{array}\right).
\end{eqnarray}
Here we immediately see that a quasideterminant solution $q_{[n+1]}$ of the Hirota equation \eqref{HE} along with its complex conjugate $r_{[n+1]}$ can be expressed as 
\begin{eqnarray}\label{qn1}
 q_{[n+1]}=q+2i\left|\begin{array}{cc} \widehat{\Theta} & e_{2n}\\  \varphi^{(n)} & \boxed{0}\end{array}\right|,\quad
 r_{[n+1]}=r+2i\left|\begin{array}{cc} \widehat{\Theta} & e_{2n-1}\\  \psi^{(n)} & \boxed{0}\end{array}\right|,
\end{eqnarray}
where it can be easily shown that the reduction $r_{[n+1]}=q_{[n+1]}^*$ holds.

\subsection{Explicit solutions}\label{ES}
In order to construct explicit solutions for the Hirota equation \eqref{HE}, we consider the quasideterminant solution given by \eqref{qn1} in which we obtain
\begin{eqnarray}\label{qHE}
q_{[n+1]}=q+2i\left|\begin{array}{cccccc}
      \varphi_1 & \psi_1^* & \ldots & \varphi_n & \psi_n^* & 0\\
      \psi_1& -\varphi_1^*& \ldots &  \psi_n & -\varphi_n^*& 0\\
    \varphi_1 \lambda_1& \psi_1^*\lambda_1^*& \ldots & \varphi_n \lambda_n & \psi_n^* \lambda_n^*& 0\\
    \psi_1 \lambda_1& -\varphi_1^*\lambda_1^*& \ldots & \psi_n \lambda_n & -\varphi_n^* \lambda_n^*& 0\\
      \vdots & \vdots &  & \vdots & \vdots & \vdots\\
      \varphi_1 \lambda_1^{n-1}& \psi_1^* \lambda_1^{*n-1} & \ldots & \varphi_n \lambda_n^{n-1} & \psi_n^* \lambda_n^{*n-1}& 0\\
      \psi_1 \lambda_1^{n-1} & -\varphi_1^*\lambda_1^{*n-1}& \ldots & \psi_n \lambda_n^{n-1} & -\varphi_n^* \lambda_n^{*n-1}& 1\\
      \varphi_1\lambda_1^n & \psi_1^*\lambda_1^{*n} &\ldots & \varphi_n \lambda_n^n & \psi_n^*\lambda_n^{*n} & \boxed{0}\\
   \end{array}
 \right|.
\end{eqnarray}
Here $\varphi_j$ and $\psi_j$ are scalar functions such that the eigenfunction $\phi_j=(\varphi_j, \psi_j)^T$ denotes $n$ distinct solutions of the spectral problem $L(\phi_j)=M(\phi_j)=0$ with the associated eigenvalue $\lambda_j$, where the operators $L$, $M$ are given by \eqref{LaxL}-\eqref{LaxM} so that
\begin{equation}\label{SEP}  
\begin{array}{ccl}
 \hspace{4cm}\phi_{j,x}+J\phi_j\lambda_j-R \phi_j&=&0,\\
\phi_{j,t}+4\beta J\phi_j\lambda_j^3+2U\phi_j\lambda_j^2 -2V\phi_j\lambda_j -W\phi_j&=&0,
 \end{array}
\end{equation}
in which $j=1,\ldots,n$ and 
$J$, $R$, $U$, $V$, $W$ are $2\times 2$ matrices given by \eqref{JRU}-\eqref{VW}. In the next section we will present  some explicit solutions of the equation \eqref{HE} for the cases $n=1,\ldots,3$. For the one-fold ($n=1$), two-fold ($n=2$) and three-fold ($n=3$) Darboux transformations, the solution \eqref{qHE} yields 
\begin{eqnarray}
 \begin{array}{ccc}\label{q2h}
  q_{[2]}=q+2i\left|\begin{array}{ccc} \varphi_1 & \psi_1^* & 0\\ \psi_1& -\varphi_1^* & 1 \\ \varphi_1\lambda_1 & \psi_1^*\lambda_1^*& \boxed{0}
  \end{array}\right|\end{array},
\end{eqnarray}
\begin{eqnarray}
 \begin{array}{ccccc}\label{q3h}
  q_{[3]}=q+2i\left|\begin{array}{ccccc} 
    \varphi_1 & \psi_1^* & \varphi_2 & \psi_2^* &  0\\ 
    \psi_1 & -\varphi_1^* & \psi_2 & -\varphi_2^* & 0 \\ 
    \varphi_1\lambda_1 & \psi_1^*\lambda_1^*& \varphi_2\lambda_2 & \psi_2^* \lambda_2^* & 0\\
     \psi_1\lambda_1 & -\varphi_1^*\lambda_1^*& \psi_2\lambda_2& -\varphi_2^*\lambda_2^* & 1\\
     \varphi_1\lambda_1^2 & \psi_1^* \lambda_1^{*2} & \varphi_2\lambda_2^2 & \psi_2^*\lambda_2^{*2} & \boxed{0}
  \end{array}\right|\end{array}
\end{eqnarray}
and
\begin{eqnarray}\label{q4h}
 \begin{array}{cccccccc}
  q_{[4]}=q+2i\left|\begin{array}{ccccccc} 
    \varphi_1 & \psi_1^* & \varphi_2 & \psi_2^* &\varphi_3&\psi_3^*&  0\\ 
    \psi_1 & -\varphi_1^* & \psi_2 & -\varphi_2^* &\psi_3&-\varphi_3^*& 0 \\ 
    \varphi_1\lambda_1 & \psi_1^*\lambda_1^*& \varphi_2\lambda_2 & \psi_2^* \lambda_2^* &  \varphi_3\lambda_3 &\psi_3^*\lambda_3^*&0\\
    \psi_1\lambda_1 & -\varphi_1^*\lambda_1^*& \psi_2\lambda_2& -\varphi_2^*\lambda_2^* &  \psi_3\lambda_3 &-\varphi_3^*\lambda_3^*&0\\
    \varphi_1\lambda_1^2 & \psi_1^*\lambda_1^{*2}& \varphi_2\lambda_2^2 & \psi_2^* \lambda_2^{*2} &  \varphi_3\lambda_3^2 &\psi_3^*\lambda_3^{*2}&0\\
    \psi_1\lambda_1^2 & -\varphi_1^*\lambda_1^{*2}& \psi_2\lambda_2^2& -\varphi_2^*\lambda_2^{*2} &  \psi_3\lambda_3^2 &-\varphi_3^*\lambda_3^{*2}&1\\ 
    \varphi_1\lambda_1^3 & \psi_1^* \lambda_1^{*3} & \varphi_2\lambda_2^3 & \psi_2^*\lambda_2^{*3} &\varphi_3\lambda_3^3&\psi_3^*\lambda_3^{*3} & \boxed{0}
  \end{array}\right|\end{array}
\end{eqnarray}
respectively. The quasideterminant solutions \eqref{q2h}-\eqref{q3h} can  be expanded as 
\begin{eqnarray}\label{q2}
 q_{[2]}=q-2i\left(\lambda_1-\lambda_1^*\right)\frac{\varphi_1 \psi_1^*}{\left|\varphi_1\right|^2+\left|\psi_1\right|^2}
\end{eqnarray}
and
\begin{eqnarray}\label{q3}
 q_{[3]}=q-2i\frac{\Lambda_{11}\left(\Pi_{12}\left|\varphi_2\right|^2+\Pi_{12}^*\left|\psi_2\right|^2\right)\varphi_1\psi_1^*+\Lambda_{22}\left(\Lambda_{12}\left|\varphi_1\right|^2+\Lambda_{12}^*\left|\psi_1\right|^2\right)\varphi_2\psi_2^*}
 {\left|\lambda_1-\lambda_2\right|^2\left|\varphi_1\varphi_2^*+\psi_1\psi_2^*\right|^2+\left|\lambda_1-\lambda_2^*\right|^2\left|\varphi_1\psi_2-\varphi_2\psi_1\right|^2},
\end{eqnarray}
where 
\begin{eqnarray*}
 \Lambda_{11}=\lambda_1-\lambda_1^*,\hspace{0.2cm} \Lambda_{22}=\lambda_2-\lambda_2^*, 
\hspace{0.2cm} \Lambda_{12}=\left(\lambda_1-\lambda_2\right)\left(\lambda_1-\lambda_2^*\right), 
\hspace{0.2cm} \Pi_{12}=\left(\lambda_1-\lambda_2\right)\left(\lambda_1^*-\lambda_2\right).
\end{eqnarray*}
Moreover, the solution \eqref{q4h} can be expressed in terms of determinants such that
\begin{eqnarray}\label{q4}
 q_{[4]}=q-2i\frac{D}{\Delta},
\end{eqnarray}
in which
\begin{eqnarray}
 \begin{array}{ccccccc}
 D&=&\left|\begin{array}{ccccccc} 
    \varphi_1 & \psi_1^* & \varphi_2 & \psi_2^* &\varphi_3&\psi_3^*\\ 
    \psi_1 & -\varphi_1^* & \psi_2 & -\varphi_2^* &\psi_3&-\varphi_3^* \\ 
    \varphi_1\lambda_1 & \psi_1^*\lambda_1^*& \varphi_2\lambda_2 & \psi_2^* \lambda_2^* &  \varphi_3\lambda_3 &\psi_3^*\lambda_3^*\\
    \psi_1\lambda_1 & -\varphi_1^*\lambda_1^*& \psi_2\lambda_2& -\varphi_2^*\lambda_2^* &  \psi_3\lambda_3 &-\varphi_3^*\lambda_3^*\\
    \varphi_1\lambda_1^2 & \psi_1^*\lambda_1^{*2}& \varphi_2\lambda_2^2 & \psi_2^* \lambda_2^{*2} &  \varphi_3\lambda_3^2 &\psi_3^*\lambda_3^{*2}\\
    \varphi_1\lambda_1^3 & \psi_1^* \lambda_1^{*3} & \varphi_2\lambda_2^3 & \psi_2^*\lambda_2^{*3} &\varphi_3\lambda_3^3&\psi_3^*\lambda_3^{*3} 
  \end{array}\right|\end{array},
\end{eqnarray}\\
\begin{eqnarray}
 \begin{array}{ccccccc}
  \Delta&=&\left|\begin{array}{cccccc} 
    \varphi_1 & \psi_1^* & \varphi_2 & \psi_2^* &\varphi_3&\psi_3^*\\ 
    \psi_1 & -\varphi_1^* & \psi_2 & -\varphi_2^* &\psi_3&-\varphi_3^* \\ 
    \varphi_1\lambda_1 & \psi_1^*\lambda_1^*& \varphi_2\lambda_2 & \psi_2^* \lambda_2^* &  \varphi_3\lambda_3 &\psi_3^*\lambda_3^*\\
    \psi_1\lambda_1 & -\varphi_1^*\lambda_1^*& \psi_2\lambda_2& -\varphi_2^*\lambda_2^* &  \psi_3\lambda_3 &-\varphi_3^*\lambda_3^*\\
    \varphi_1\lambda_1^2 & \psi_1^*\lambda_1^{*2}& \varphi_2\lambda_2^2 & \psi_2^* \lambda_2^{*2} &  \varphi_3\lambda_3^2 &\psi_3^*\lambda_3^{*2}\\
    \psi_1\lambda_1^2 & -\varphi_1^*\lambda_1^{*2}& \psi_2\lambda_2^2& -\varphi_2^*\lambda_2^{*2} &  \psi_3\lambda_3^2 &-\varphi_3^*\lambda_3^{*2}\\ 
  \end{array}\right|\end{array}.
\end{eqnarray}

\section{Particular solutions}\label{ParS}

\subsection{Solutions for zero seed}
For $q=r=0$, the spectral problem \eqref{SEP} becomes
\begin{equation}
 \begin{array}{ccl}
   \hspace{2.2cm} \phi_{j,x}+J\phi_j\lambda_j&=&0,\\
   \phi_{j,t}+\left(4\beta\lambda_j^3+2\alpha \lambda_j^2\right)J\phi_j&=&0,
\end{array}
\end{equation}
which has solution $\phi_j=\left(\varphi_j, \psi_j\right)^T$ such that
\begin{equation}
 \begin{array}{ccl}\label{varphipsi}
    \varphi_j\left(x,t,\lambda_j\right)&=&e^{-i\left[\lambda_j x+\left(2\alpha\lambda_j^2+4\beta\lambda_j^3\right)t\right]},\\
        \psi_j\left(x,t,\lambda_j\right)&=&e^{i\left[\lambda_j x+\left(2\alpha\lambda_j^2+4\beta\lambda_j^3\right)t\right]},
\end{array}
\end{equation}
where $j=1,\ldots,n$.

\subsubsection*{Case I (${n=1}$)}
\begin{figure}[ht!]
    \centering
 \subfloat[($a$)]{{\includegraphics[width=7.7cm]{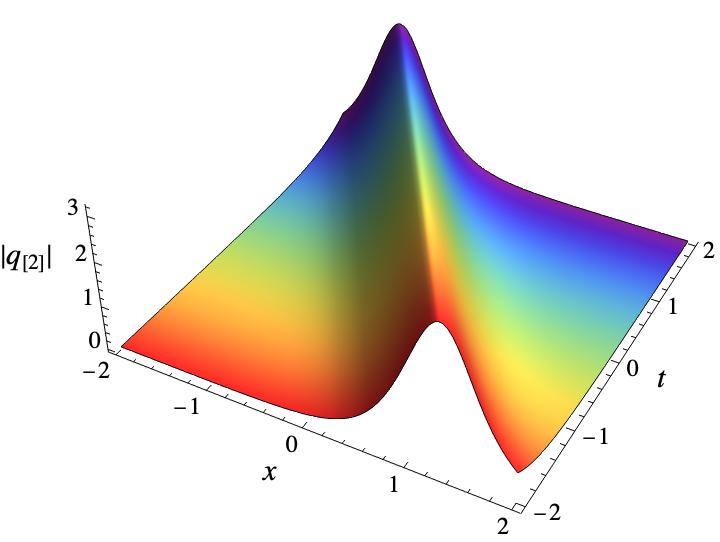} }}
    \qquad
    \subfloat[($b$)]{{\includegraphics[width=7.7cm]{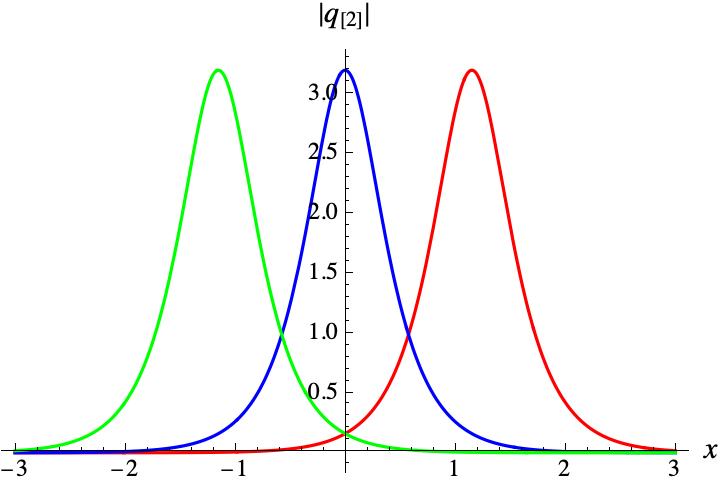} }}
    \caption{(Color online) One-soliton solution $|q_{[2]}|$ of the HE \eqref{HE} when $\alpha=\beta=1$, $\xi=0.8, \eta=1.6$. Figure ($a$) describes its surface and ($b$) gives its profiles at different times $t=-1.8$ (red), $t=0$ (blue), $t=1.8$ (green).}
     \label{fig1}
\end{figure}

By letting $\lambda_1=\xi+i\eta$ and substituting the functions $\varphi_1$ and $\psi_1$ given  by \eqref{varphipsi} into \eqref{q2}, we obtain the one-soliton solution of the Hirota equation \eqref{HE} as
\begin{eqnarray}
 q_{[2]}=2\eta e^{-2i\left[\xi x+2\left(\alpha\left[\xi^2-\eta^2\right]+2\beta\left[\xi^3-3\xi\eta^2\right]\right)t\right]} 
 \sech\left(2\eta x+8\left[\alpha\xi\eta+\beta\left(3\xi^2\eta-\eta^3\right)t\right]\right)
\end{eqnarray}
which yields 
\begin{eqnarray}
 \left|q_{[2]}\right|^2=4\eta^2 \sech^2\left(2\eta x+8\left[\alpha\xi\eta+\beta\left(3\xi^2\eta-\eta^3\right)\right]t\right).
\end{eqnarray}
This solution is plotted in \autoref{fig1}.
\subsubsection*{Case II (${n=2}$)}
\begin{figure}[ht!]
    \centering
    \subfloat[($a$)]{{\includegraphics[width=8.6cm]{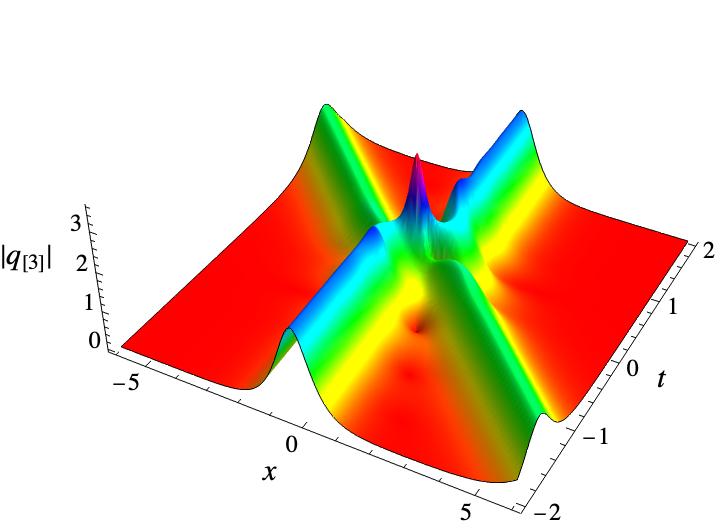} }}
    \qquad
    \subfloat[($b$)]{{\includegraphics[width=6.8cm]{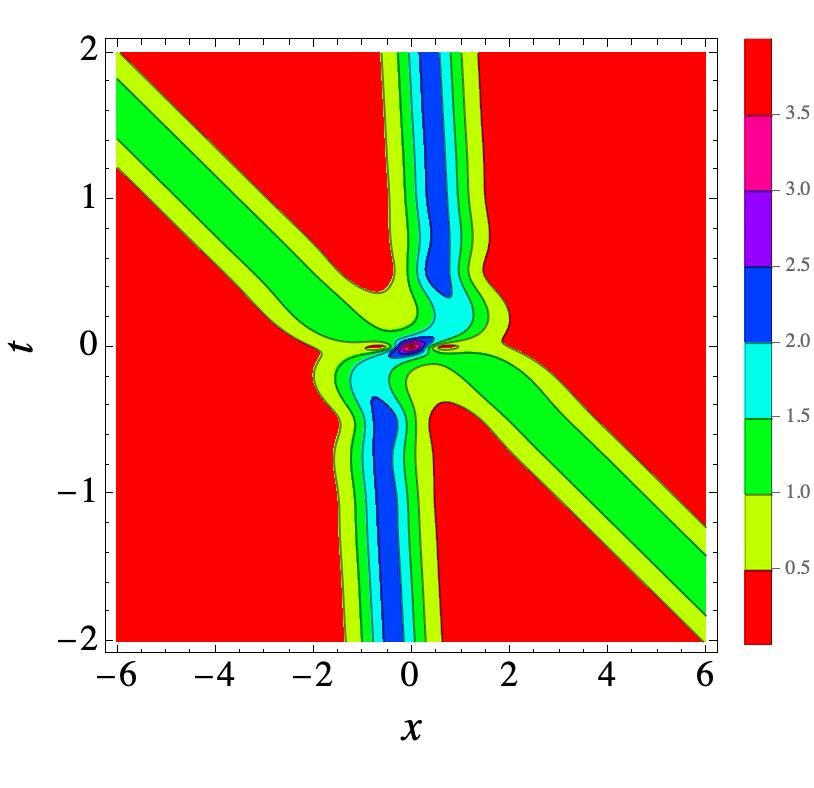} }}
    \caption{(Color online) Two-soliton solution $|q_{[3]}|$ of the HE \eqref{HE} when $\alpha=\beta=1$, $\xi=0.5$, $\eta_1=0.7$ and $\eta_2=1.1$. ($a$) Surface diagram. ($b$) Contour diagram.}
     \label{fig2}
\end{figure}

 Let $\lambda_1=\xi+\eta_1$ and $\lambda_2=\xi+\eta_2$ such that $\eta_1\eta_2\neq 0$. By substituting the corresponding eigenfunctions $\varphi_1, \psi_1$ and $\varphi_2, \psi_2$, given by \eqref{varphipsi}, into \eqref{q3}, we obtain the two-soliton solution of the Hirota equation \eqref{HE} as
\begin{eqnarray}
  q_{[3]}=4\left(\eta_1^2-\eta_2^2\right)\frac{\eta_1 e^{-ig_1}\cosh f_2-\eta_2 e^{-ig_2} \cosh f_1}{\left(\eta_1-\eta_2\right)^2\cosh F_1+\left(\eta_1+\eta_2\right)^2\cosh F_2-4\eta_1\eta_2\cos F_3}
\end{eqnarray}
which yields
\begin{eqnarray}
  \left|q_{[3]}\right|^2=16\left(\eta_1^2-\eta_2^2\right)^2\frac{\eta_2^2\cosh^2 f_1+\eta_1^2\cosh^2f_2-2\eta_1\eta_2\cosh f_1\cosh f_2\cos F_3 } {\left[\left(\eta_1-\eta_2\right)^2\cosh F_1+\left(\eta_1+\eta_2\right)^2\cosh F_2-4\eta_1\eta_2\cos F_3\right]^2}  ,
\end{eqnarray}
where
\begin{eqnarray*}
  f_1&=&2\eta_1\left[x+4\left(\alpha\xi+\beta\left[3\xi^2-\eta_1^2\right]\right)t\right],\\
  f_2&=&2\eta_2\left[x+4\left(\alpha\xi+\beta\left[3\xi^2-\eta_2^2\right]\right)t\right],\\
  g_1&=&2\xi x+4\left[\alpha\left(\xi^2-\eta_1^2\right)+2\beta\xi\left(\xi^2-3\eta_1^2\right)\right]t,\\
  g_2&=&2\xi x+4\left[\alpha\left(\xi^2-\eta_2^2\right)+2\beta\xi\left(\xi^2-3\eta_2^2\right)\right]t
\end{eqnarray*}
and $F_1=f_1+f_2$, $F_2=f_1-f_2$, $F_3=g_1-g_2$ such that
\begin{eqnarray*}
  F_1&=&2\left(\eta_1+\eta_2\right)\left[x+4\left(\alpha\xi+\beta\left[3\xi^2+\eta_1\eta_2-\eta_1^2-\eta_2^2\right]\right)t\right],\\
  F_2&=&2\left(\eta_1-\eta_2\right)\left[x+4\left(\alpha\xi+\beta\left[3\xi^2-\eta_1\eta_2-\eta_1^2-\eta_2^2\right]\right)t\right],\\
  F_3&=&4\left(\eta_2^2-\eta_1^2\right)\left[\alpha+6\beta\xi\right]t.
\end{eqnarray*}
By choosing appropriate parameters, the two-soliton solution of the Hirota equation \eqref{HE} is plotted in \autoref{fig2}.

\subsubsection*{Case III (${n=3}$)}

\begin{figure}[ht!]
    \centering
    \subfloat[($a$)]{{\includegraphics[width=8.6cm]{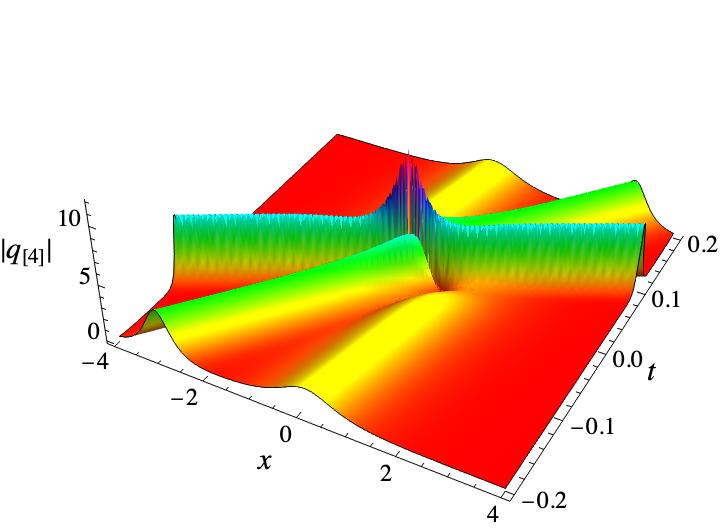} }}
    \qquad
    \subfloat[($b$)]{{\includegraphics[width=6.8cm]{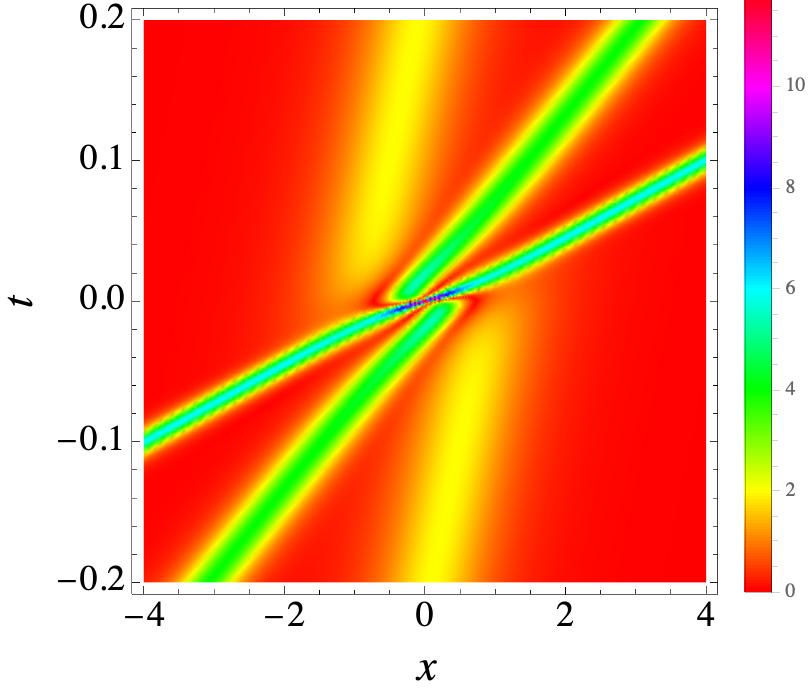} }}
    \caption{(Color online) Three-soliton solution $|q_{[4]}|$ of the HE \eqref{HE} when $\lambda_1=i$, $\lambda_2=2i$ and $\lambda_3=3i$. ($a$) Surface diagram. ($b$) Density diagram.}
     \label{fig3}
\end{figure}

In this case, we have three eigenvalues $\lambda_1$, $\lambda_2$ and $\lambda_3$. Let us choose $\lambda_1=i$, $\lambda_2=2i$ and $\lambda_3=3i$. By substituting the corresponding eigenfunctions $(\varphi_1, \psi_1)^T$, $(\varphi_2, \psi_2)^T$ and $(\varphi_3, \psi_3)^T$, given by \eqref{varphipsi}, into \eqref{q4}, we obtain the three-soliton solution of the Hirota equation \eqref{HE}. By choosing appropriate parameters, this solution  is plotted in \autoref{fig3}.

\subsection{Solutions for non-zero seed}
In this subsection, for $q, r\neq 0$ and $r=q^*$, we take $q=c e^{i\mu}$ as a plane wave solution of the Hirota equation \eqref{HE}, where $\mu=ax+bt$ in which $a, b, c \in \mathbb{R}$ under the condition $b=\alpha\left(2c^2-a^2\right)+\beta\left(a^3-6ac^2\right)$. We use this as a seed solution. Substituting $q=c e^{i\mu}$ into the linear system \eqref{SEP} and then solving for the eigenfunction $\phi_j=\left(\varphi_j, \psi_j\right)^T$, we obtain
\begin{equation}
 \begin{array}{ccl}\label{efj}
   \varphi_j\left(x,t,\lambda_j\right)&=&e^{\frac{1}{2}i\mu}\left(c_j e^{\frac{1}{2}i\gamma_j}+e_j e^{-\frac{1}{2}i\gamma_j}\right),\\
   \psi_j\left(x,t,\lambda_j\right)&=&e^{-\frac{1}{2}i\mu}\left(\widetilde{c_j} e^{\frac{1}{2}i\gamma_j}+\widetilde{e_j} e^{-\frac{1}{2}i\gamma_j}\right),
\end{array}
\end{equation}
where 
\begin{eqnarray*}
    \gamma_j&=&s_j\left(x+k_jt\right), \\
    \widetilde{c_j}&=&i\frac{c_j}{2c}\left(a+2\lambda_j+s_j\right),\\
    \widetilde{e_j}&=&i\frac{e_j}{2c}\left(a+2\lambda_j-s_j\right)
\end{eqnarray*}
in which $s_j=\sqrt{\left(a+2\lambda_j\right)^2+4c^2}$, $k_j=\alpha\left(2\lambda_j-a\right)+\beta\left(a^2-2a\lambda_j+4\lambda_j^2-2c^2\right)$ and $c_j$, $e_j$ are arbitrary constants such that $j=1,\ldots,n$.

\subsubsection*{Case IV (${n=1}$)}
\begin{figure}
    \centering
    \subfloat[($a$)]{{\includegraphics[width=8.6cm]{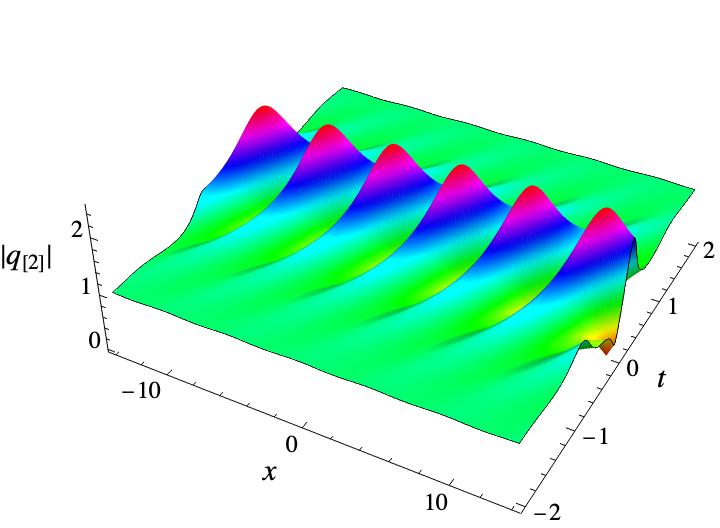} }}
    \qquad
    \subfloat[($b$)]{{\includegraphics[width=6.8cm]{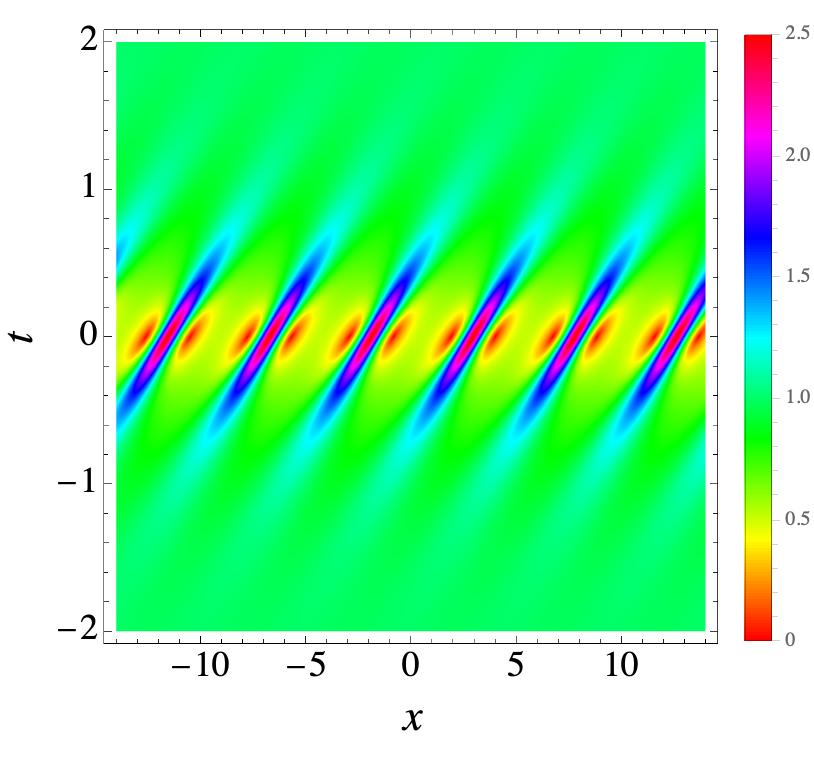} }}
    \caption{(Color online) Breather solution $|q_{[2]}|$ of the HE \eqref{HE} when $\alpha=\beta=1$, $\xi=0.04$ and $\eta=0.76$. ($a$) Surface diagram. ($b$) Density diagram.}
     \label{fig4}
\end{figure}

Let the eigenvalue $\lambda_1=\xi+i\eta$. For simplicity, choose $a=-2\xi$ and $c_1=e_1=c$. Substituting the seed solution $q=c e^{i\mu}$ and the functions $\varphi_1$, $\psi_1$ given by \eqref{efj}, into \eqref{q2}, we obtain the following breather solution 
\begin{eqnarray}
  q_{[2]}=c e^{i\mu}\left(1-2\eta\frac{\eta\cosh (\Omega t)-i\omega\sinh (\Omega t)+\eta\cos [2\omega (x+\Gamma t)]+\omega\sin [2\omega (x+\Gamma t)]}
    {c^2\cosh (\Omega t)+\eta^2\cos  [2\omega (x+\Gamma t)]+\eta\omega\sin  [2\omega (x+\Gamma t)]}\right),
\end{eqnarray}
where
\begin{eqnarray*}
  \mu&=&-2\xi x+\left[2\alpha\left(c^2-2\xi^2\right)+4\beta\left(3c^2\xi-2\xi^3\right)\right]t,\\
  \omega&=&\sqrt{c^2-\eta^2},\\
  \Omega&=&4\eta\omega (\alpha+6\beta \xi),\\
  \Gamma&=&4\alpha\xi+2\beta\left(6\xi^2-2\eta^2-c^2\right).
\end{eqnarray*}
Thus, we have
\begin{eqnarray}
   \left|q_{[2]}\right|^2=c^2 \frac{F^2+G^2}{H^2},
\end{eqnarray}
where
\begin{eqnarray*}
  F&=&\left(2\eta^2-c^2\right) \cosh (\Omega t)+\eta^2 \cos [2w (x+\Gamma t)]+\eta w \sin [2w (x+\Gamma t)],\\
  G&=&2\eta w \sinh (\Omega t)\\
  H&=&c^2 \cosh (\Omega t)+\eta^2 \cos [2w (x+\Gamma t)]+\eta w \sin [2w (x+\Gamma t)].
\end{eqnarray*}
\autoref{fig4} shows the dynamical evolution of the breather solution of the Hirota equation \eqref{HE}.

\section{Conclusion}\label{Con}
In conclusion, we have studied a standard Darboux transformation to construct quasideterminant solutions for the Hirota equation \eqref{HE}. These quasideterminants are expressed in terms of  solutions of  the linear partial differential equations given by  \eqref{SEP}. It should be highlighted that these quasideterminant solutions arise naturally from the Darboux transformation we present here.
Furthermore, the multisoliton and breather solutions for zero and non-zero seeds have been given as particular examples for the HE.  Examples of these particular solutions are plotted in the figures $\ref{fig1}-\ref{fig4}$ with the chosen parameters. Finally, we point out that the method we have presented in this paper allows us to construct exact solutions for other integrable nonlinear evolution equations  such as \cite{GM,HY,NY14}.


\end{document}